\documentclass[aps,prb,superscriptaddress,longbibliography,twocolumn]{revtex4-2}
\usepackage{bm}
\usepackage{lmodern}

\usepackage[utf8]{inputenc}
\usepackage{amsmath}
\usepackage{amssymb}
\usepackage{graphicx}
\usepackage{epstopdf}
\usepackage{xcolor}
\usepackage{enumerate}
\usepackage[normalem]{ulem}
\usepackage[english]{babel}
\usepackage{braket}
\usepackage{natbib}
\usepackage{float}
\usepackage{stmaryrd}
\usepackage{hyperref}
\usepackage{physics}
\usepackage{upgreek}
\usepackage[T1]{fontenc}
\usepackage{textcomp}

\begin{document}

\title{Electrical control of uniformity in quantum dot devices}

\author{Marcel Meyer}
\author{Corentin D\'{e}prez}
\author{Timo R. van Abswoude}
\author{Dingshan Liu}
\author{Chien-An Wang}
\affiliation{QuTech and Kavli Institute of Nanoscience, Delft University of Technology, PO Box 5046, 2600 GA Delft, The Netherlands}
\author{Saurabh Karwal}
\author{Stefan Oosterhout}
\affiliation{QuTech and Netherlands Organisation for Applied Scientific Research (TNO), Delft, The Netherlands}
\author{Francesco Borsoi}
\affiliation{QuTech and Kavli Institute of Nanoscience, Delft University of Technology, PO Box 5046, 2600 GA Delft, The Netherlands}
\author{Amir Sammak}
\affiliation{QuTech and Netherlands Organisation for Applied Scientific Research (TNO), Delft, The Netherlands}
\author{Nico W. Hendrickx}
\author{Giordano Scappucci}
\author{Menno Veldhorst}
\affiliation{QuTech and Kavli Institute of Nanoscience, Delft University of Technology, PO Box 5046, 2600 GA Delft, The Netherlands}

\date{\today}

\begin{abstract}
Highly uniform quantum systems are essential for the practical implementation of scalable quantum processors. While quantum dot spin qubits based on semiconductor technology are a promising platform for large-scale quantum computing, their small size makes them particularly sensitive to their local environment. Here, we present a method to electrically obtain a high degree of uniformity in the intrinsic potential landscape using hysteretic shifts of the gate voltage characteristics. We demonstrate the tuning of pinch-off voltages in quantum dot devices  over hundreds of millivolts that then remain stable at least for hours. Applying our method, we homogenize the pinch-off voltages of the plunger gates in a linear array for four quantum dots reducing the spread in pinch-off voltage by one order of magnitude. This work provides a new tool for the tuning of quantum dot devices and offers new perspectives for the implementation of scalable spin qubit arrays.
\end{abstract}

\maketitle

Spin qubits in semiconductor quantum dots are a promising platform for quantum information processing~\cite{Zwanenburg2013, vandersypen2017,Scappucci2020, Scappucci2021}. Group IV semiconductors such as silicon and germanium can be isotopically purified~\cite{Itoh2014}, enabling long quantum coherence~\cite{Veldhorst2014, Stano2022}, high-fidelity single-qubit~\cite{Dehollain2016,Yoneda2018,Lawrie2021} and two-qubit gates~\cite{Madzik2022,Noiri2022,Xue2022} as well as multi-qubit operation~\cite{Hendrickx2021,Philips2022}. Spin qubits can be operated at comparatively high temperatures~\cite{Petit2020,Yang2020,Camenzind2022} and their compatibility with semiconductor technologies spurred the realization of qubits made in industrial foundries~\cite{Ansaloni2020,Zwerver2022}. However, implementing more than a few qubits on a single chip remains extremely challenging.

Variations, in particular at the nanoscale, may lead to significant alterations of the relevant device metrics~\cite{Zwanenburg2013, vandersypen2017,Bavdaz2022}, such as the voltage needed to load a single electron to be used as a spin qubit. These variations can complicate the tuning of initialization, control or readout and potentially form a roadblock for larger systems. Additionally, qubit-to-qubit variability may require the use of individual control electronics for each qubit, as is common practice in current experimental implementations, thus challenging the scalability. While several proposals have been put forward to scale quantum dot qubits~\cite{Hill2015,vandersypen2017,Veldhorst2017,Li2018}, in all cases a high level of device uniformity is critical in their realization.

\begin{figure*}[htb]
\centering
\includegraphics[width=0.84\textwidth]{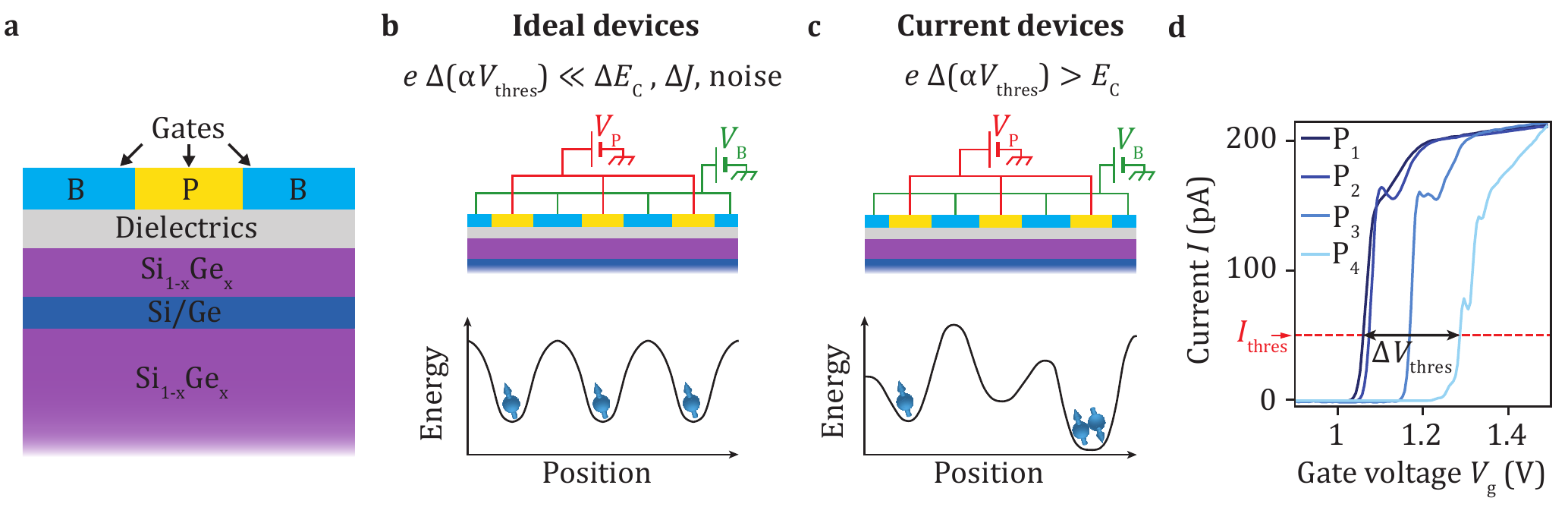}
\caption{\textbf{Fluctuations in the potential landscape in semiconductor quantum dot devices. a}, Schematics of typical semiconductor heterostructures with buried quantum wells studied. The metallic gate electrodes colored in blue and yellow represent the barrier (B) and plunger gates (P) of a quantum dot array, respectively. \textbf{b}, Potential landscape in an ideal device with shared gate control. The application of the same voltage $V_{\rm P/B}$ on all plunger/barrier gates leads to a regular potential landscape with fluctuations negligible compared to those of the other relevant energy scales ($\alpha$ denotes the gate lever arm). The quantum dots all have the same charge configuration. \textbf{c}, Potential landscape in state of the art devices with shared gate control. The application of the same voltage $V_{\rm P/B}$ on all plunger/barrier gates leads to an irregular potential landscape due to local fluctuations which are often comparable or larger than the charging energy $E_{\rm C}$. Consequently, the quantum dots have different charge configurations. \textbf{d}, Typical variations in the pinch-off characteristic of the plunger gates in a state-of-the-art linear quantum dot array (device~A), nominally identical to the one displayed in in Fig.~\ref{fig:Fig4}.a, just after a cooldown. The pinch-off voltage $V_{\rm thres}$ is defined as the gate voltage for which the current reaches $I_{\rm thres}=50~$pA at a bias of $\abs{V_{\rm sd}}=100~\upmu$V. Here, the pinch-off voltages spread over a voltage range $\Delta V_{\text{ thres}}=225~$mV. }
\label{fig:Fig1}
\end{figure*}

For semiconductor quantum dot qubits, the uniformity of the potential landscape is the key parameter that dictates the number of control voltages required per qubit. Ideally, a few voltages would suffice to induce a highly regular potential landscape as drawn in Fig.~\ref{fig:Fig1}.b. Yet, potential fluctuations are naturally present as illustrated in Fig.~\ref{fig:Fig1}.c. They can be caused by defects and charge traps, mechanical stress induced by the deposition of metallic gates~\cite{Thorbeck2015,Stein2021}, as well as variations in material growth or in the exact shape of the gates. The development of devices based on quantum wells buried in heterostructures, similar to that sketched in Fig.~\ref{fig:Fig1}.a, already has led to a drastic improvement of the uniformity compared to metal-oxide-semiconductor systems~\cite{Lawrie2020}. This has enabled the control of up to 16 quantum dots in a four-by-four array with shared gate control~\cite{Borsoi2022}. However, significant variations in the quantum dot potential landscape are still commonly observed~\cite{Zajac2016,Mills2019,Borsoi2022}. This raises the question whether material~\cite{Scappucci2021} and fabrication development~\cite{Dodson2020,Ha2022,Zwerver2022,Borsoi2022} will suffice to reach the required uniformity to operate large qubit arrays.

Here, we present an alternative method and demonstrate electrical control of quantum dot uniformity. Our approach takes advantage of the gate voltage hysteresis, an ubiquitous effect observed in semiconductor heterostructures, that is mostly considered as a limitation in the tune-up of quantum dots. It manifests in shifts of the gate voltage characteristics and is commonly explained by a build up of charges at the interface between the semiconductor barrier and the dielectrics which then alter the electric field in the buried quantum well~\cite{Lu2011,Huang2014,Laroche2015,Su2017,Chou2018,Su2019,Esposti2022}. We unveil the hysteresis and its effects on the potential landscape beneath the gates by studying how pinch-off characteristics evolve with the application of tailored stress voltage sequences. This method allows us to tune those pinch-off voltages over hundreds of millivolts after which they remain stable at least on the time scale of hours. We then apply our findings to homogeneize the plunger gate pinch-off characteristics in a linear quantum dot array reducing potential fluctuations in the quantum well underneath the corresponding gates.

The gate voltage required to confine a single electron or hole typically varies between quantum dots in an array as it is dependent on the local electrostatic environment. These fluctuations also affect the pinch-off curve as exemplary depicted for sweeping the four plunger gates of a linear quantum dot device (similar to that shown in Fig~\ref{fig:Fig4}.a) in Fig.~\ref{fig:Fig1}.d. The curves reveal the local depletion of a conducting path through the quantum well and experimentally can be obtained in a very short time compared to the time required for the formation of a well defined quantum dot. Therefore, we will employ pinch-off characteristics in the following to efficiently estimate variations in the potential landscape on the length scale of single quantum dots. In particular, we focus on the pinch-off voltages $V_{\rm thres}$ defined as the gate voltages at which a current of $I_{\rm thres}=$~50~pA is reached for an applied source drain bias of $\abs{V_{\rm sd}}=100~\upmu$V.

\begin{figure*}[htb]
\centering
\includegraphics[width=0.84\textwidth]{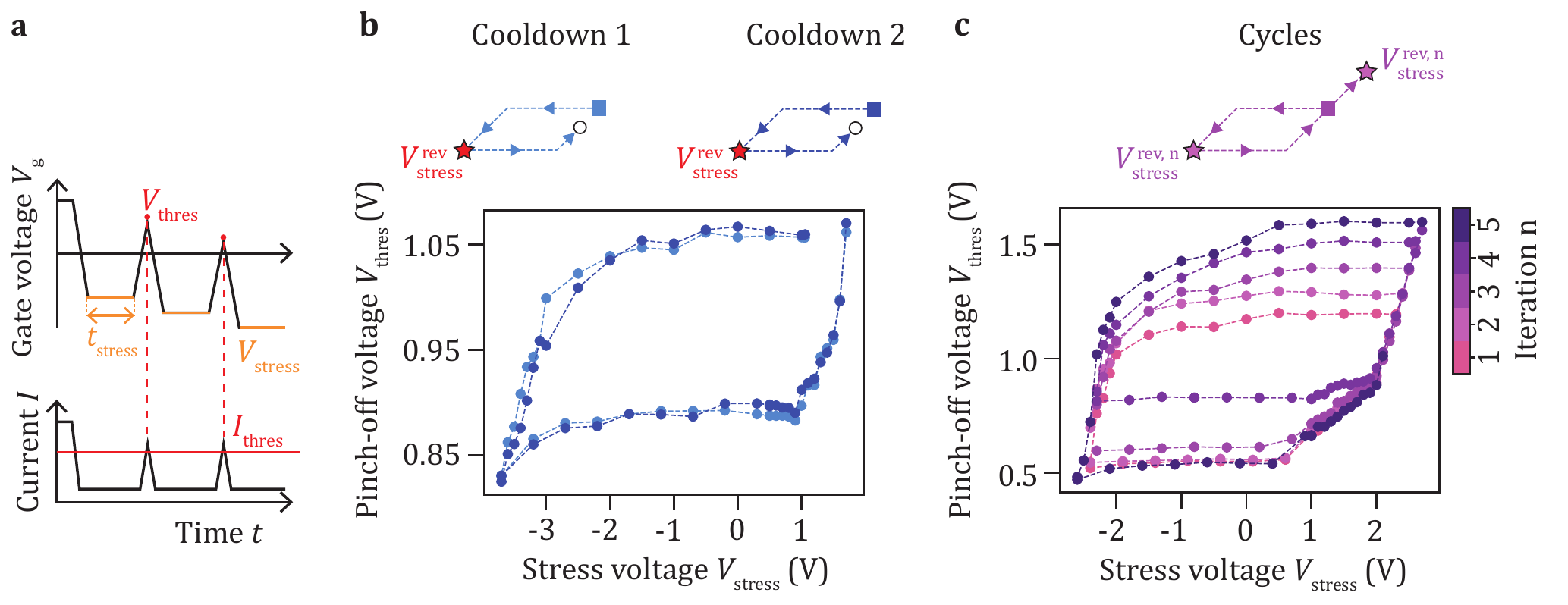}
\caption{\textbf{Hysteresis of the pinch-off characteristics. a}, Schematics of the measurement sequence used to probe the hysteretic behavior of the pinch-off voltage $V_{\rm thres}$ of a single gate. $V_{\rm thres}$, i.e. the voltage when the current reaches $I_{\rm thres}=50~$pA at a bias voltage $\abs{V_{\rm sd}}=100~\upmu$V, is measured  after application of successive stress voltages $V_{\rm stress}$ for $t_{\rm stress}=1$~min. \textbf{b}, Evolution of the pinch-off voltage $V_{\rm thres}$ of the sensor plunger gate~S as a function of the stress voltage $V_{\rm stress}$ for two different cooldowns of device~B. The measurement cycle is sketched in the top illustration. The square and the circle mark the starting point and the ending point of the cycles, respectively. The star indicates the point $V_{\rm stress}^{\rm rev}$ where the stress voltage sequence is reversed. $V_{\rm stress}$ is first  decreased before being increased again after $V_{\rm stress}=3.7~$V. Both sets of points draw hysteresis cycles which overlap. The remaining gates that are needed to form a conductive channel are set to $V_{\rm 0}=1.2~$V. \textbf{c}, $(V_{\rm stress}$,$V_{\rm thres})$ hysteresis cycles measured successively for plunger gate $\rm P_{\rm 1}$ in device~A. The points where the stress voltage sequences are reversed (star) and ended (circle) are changed between each cycle. Note that for the fifth iteration, the stress voltage sequence with increasing $V_{\rm stress}$ was stopped purposely when $V_{\rm thres}\simeq 1~$V. All other gates are set to $V_{\rm 0}=1.704~$V.}
\label{fig:Fig2}
\end{figure*}

We study devices in $^{28}$Si/SiGe heterostructures~\cite{Paquelet2021} and investigate how the pinch-off voltage of a single gate evolves depending on the previously applied gate voltages. To that end, we conduct systematic transport measurements at 4.2~K similar to sequences in~\cite{Ershov2003, Kaczer2008, Lelis2015, Franco2014} following the procedure depicted in Fig.~\ref{fig:Fig2}.a. First a stress voltage $V_{\rm stress}$ is applied to the gate under study for a time $t_{\rm stress}=1$~min. Then the gate-voltage is swept back until the pinch-off condition $I=I_{\rm thres}$ is met. This sequence is repeated several times with evolving stress voltages to measure the evolution of $V_{\rm thres}$ as a function of $V_{\rm stress}$. First, the applied stress voltage $V_{\rm stress}$ is decreased step-wise to be increased gradually again (not illustrated) after reaching a reversal point $V_{\rm stress}=V_{\rm stress}^{\rm rev}$.

Fig.~\ref{fig:Fig2}.b shows the resulting pinch-off voltage evolution for a plunger gate $\rm P_{\rm i}$ that is part of a linear quantum dot array for two different cooldowns (light blue and dark blue curve, respectively). In these cases, $V_{\rm stress}$ is first lowered step-wise from $V_{\rm stress}=1.05$~V to $V_{\rm stress}^{\rm rev}=-3.7~$V. We observe that up to $V_{\rm stress}>-2.0$~V the pinch-off voltages $V_{\rm thres}$ stay within $\pm 15$~mV of the first pinch-off voltage $V_{\rm thres}^0=1.06$~V forming a plateau. Then, they drop down rapidly to $V_{\rm thres}=0.83~$V. At $V_{\rm stress}^{\rm rev}=-3.7~$V, the sweep direction is reversed and we start to increase $V_{\rm stress}$ progressively. However, we do not observe a reversed behavior. Instead, from $V_{\rm stress}=-2.7$~V to $V_{\rm stress}=0.9$~V, the pinch-off voltages increase by less than $25~$mV forming a second plateau. Only when $V_{\rm stress}=1.0(1.1)~\rm{V}$ for the first(second) cooldown, $V_{\rm thres}$ starts to increase steeply again. The ensembles of ($V_{\rm stress}$,$V_{\rm thres}$) values draw typical hysteresis cycles with plateaus marking the ranges of applicable gate voltages over which the pinch-off voltage is not significantly changing. Furthermore, Fig.~\ref{fig:Fig2}.b highlights the effect of thermal cycling on these measurements and reveals a remarkable overlap of the hysteresis cycles measured during two different cooldowns. A high degree of similarity is also observed when comparing successive measurements performed using the same stress voltage sequence as shown in supplementary Fig.~S2 for gate~S of device~D. This suggests that the underlying process has a deterministic nature.

Similar experiments performed on another sample with varying reversal points $V_{\rm stress}^{\rm rev}$ result in the cycles plotted in Fig.~\ref{fig:Fig2}.c. The shape of the curves is nearly identical for each iteration. Again, we observe plateaus where the pinch-off voltage deviates by less than $50~$mV from its first value. Yet, the position of the plateaus varies with the chosen $V_{\rm stress}^{\rm rev}$. The pinch-off voltage plateaus can be shifted by up to $\abs{\Delta V_{\rm thres}}=290$~mV for the lower plateau and by up to $\abs{\Delta V_{\rm thres}}=400$~mV for the upper one. Overall, Fig.~\ref{fig:Fig2}.b and Fig.~\ref{fig:Fig2}.c suggest that by applying a dedicated voltage sequence the pinch-off voltage can be adjusted on-demand to chosen targets and thus that the intrinsic potential landscape underneath the gates can be tuned.

We also note that similar hysteretic behaviours, with sample-dependent variations of the exact shape of the ($V_{\rm stress}$,$V_{\rm thres}$) curves, are consistently found in several Si/SiGe devices (e.g. device~D gate~S shown in supplementary Fig.~S2) as well as in samples made from Ge/SiGe heterostructures (see supplementary Fig.~S3) suggesting a common underlying mechanism. The observed reproducibility and the large control window of the pinch-off voltage are the foundations of our approach to homogenize the potential landscape below an ensemble of gates.

\begin{figure}[htb]
\centering
\includegraphics[width=0.84\linewidth]{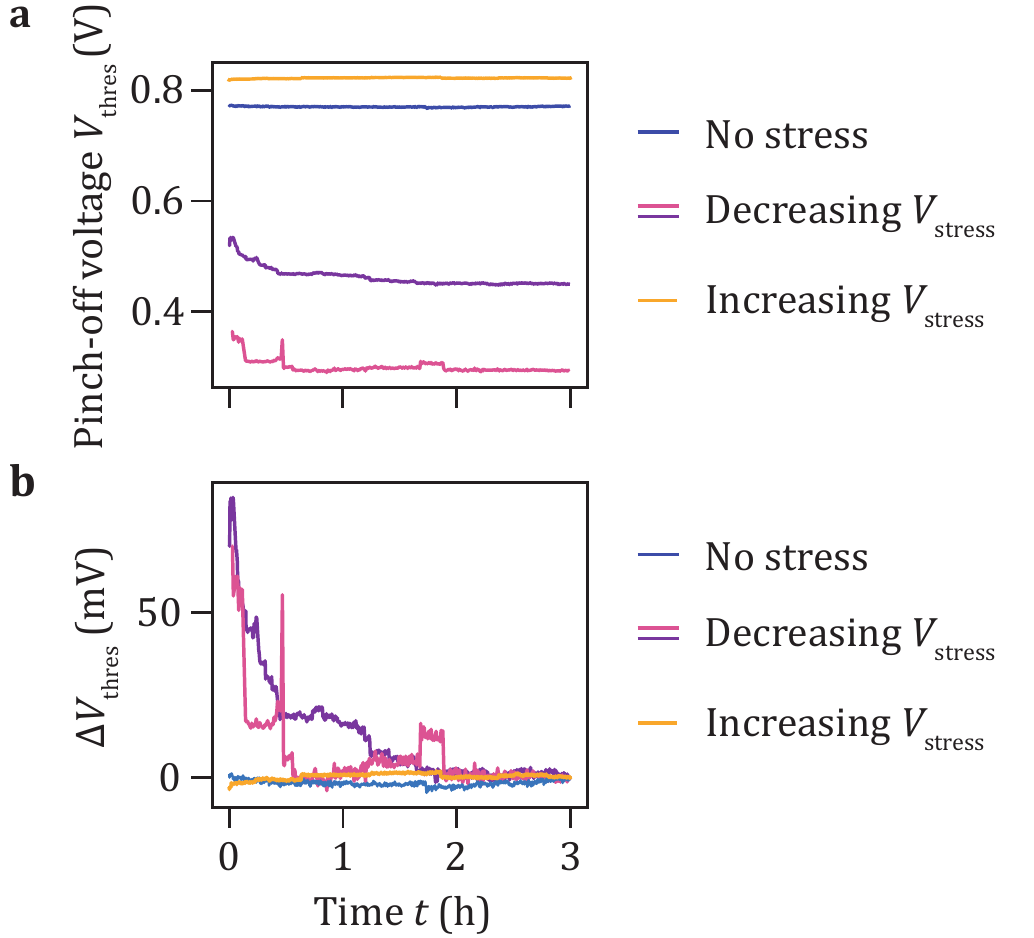}
\caption{\textbf{Stability of the pinch-off voltage after tuning. a}, Time evolution of the $V_{\rm thres}$ prior to any application of stress voltages blue) and after tuning via application of increasing $V_{\rm stress}$ with $V_{\rm stress}^{\rm rev}>0~$V (orange) or decreasing $V_{\rm stress}$ with $V_{\rm stress}^{\rm rev}<0~$V (pink and violet). The curves are obtained for sensor plunger gate~S in device~C, except of the pink curve which is obtained for sensor plunger gate~S in device~D. \textbf{b}, Relative variations $\Delta V_{\rm thres}= V_{\rm thres}(t)-V_{\rm thres}(t=3~\text{h})$ of the data shown in \textbf{a}. }
\label{fig:Fig3}
\end{figure}

However, the electrical tuning of the intrinsic potential uniformity is of practical interest only if the resulting potential landscape remains stable afterwards. Therefore, we study how the pinch-off voltage evolves in time after stopping the hysteresis measurement cycle at varying points (see supplementary section~I.C for the detailed experimental procedure). Fig.~\ref{fig:Fig3}.a shows the time evolution directly after the application of decreasing (pink/violet) and increasing (orange) stress voltages. For comparison, we also plot how the pinch-off voltage evolves right after a cooldown without prior application of a stress voltage sequence (blue). For decreasing $V_{\rm stress}$ sequences, the pinch-off voltages converge into steady states after initial decays and the time evolution exhibits random abrupt jumps. For the situation where no stress voltage or increasing stress voltages $V_{\rm stress}$ are applied no significant variations of $V_{\rm thres}$ are observed. The relative evolution depicted in Fig.~\ref{fig:Fig3}.b reveals that for $t>2~$hours, the voltage fluctuations are similar for all three situations. This is confirmed by extracting the standard deviations of $V_{\rm thres}$ for experiments with and without application of stress voltage sequence which are $\sigma_{\rm stress}=0.4$~mV (orange),  $\sigma_{\rm stress}=1.0$~mV (pink and violet) and $\sigma_{\rm no~stress}=0.8$~mV (blue), respectively. These experiments suggest that after a potential initial transient regime there is no change in the stability of the device due to the electrical tuning. This stability is observed for at least one hour and up to three depending on the voltage sequence applied.

\begin{figure*}[htb]
\centering
\includegraphics[width=0.84\textwidth]{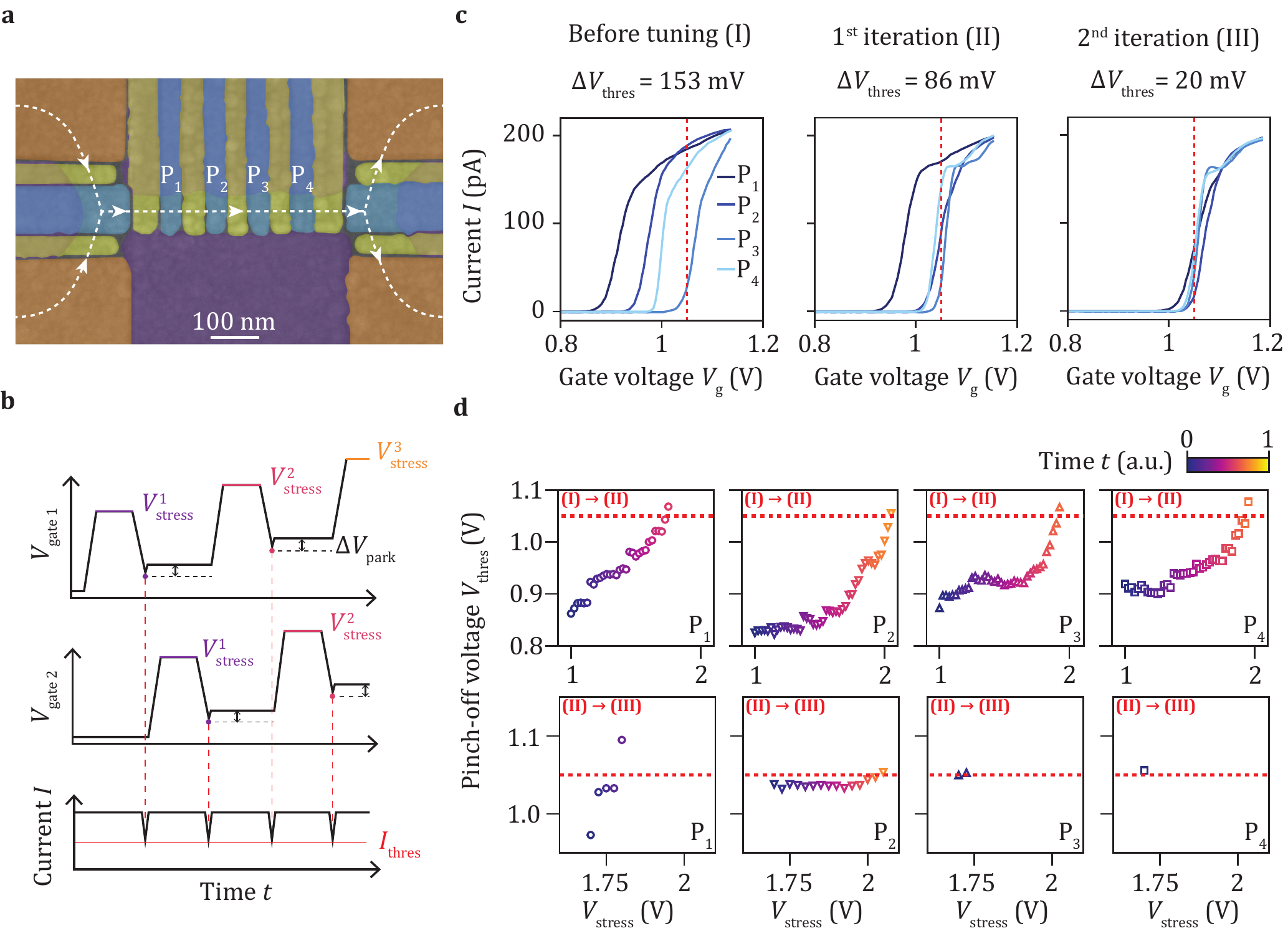}
\caption{\textbf{Homogenization of the potential landscape below the plunger gates of a linear quantum dot array. a}, Scanning electron micrograph of a linear quantum dot array. The plunger, barrier, accumulation and screening gates are colored in blue, yellow, orange and violet, respectively. The current flow is depicted by the dashed line. We aim at equalizing the pinch-off voltages of the plunger gates $\rm P_{\rm i}$. \textbf{b}, Schematics of the strategy followed illustrated with only two gates for clarity. Note that here, in contrast to the illustration in Fig.~\ref{fig:Fig2}.A, the pinch-off voltage $V_{\rm thres}$ is detected through lowering the gate voltage until $I=I_{\rm thres}$. \textbf{c}, Evolution of the pinch-off characteristics in device~A after two iterations of the tuning procedure. The target voltage $V_{\rm target}=1.05~$V is marked by a red dashed line. After two iterations the spread of the pinch-off voltage $\Delta V_{\rm thres}$ is reduced from 153~mV to 20 mV. \textbf{d}, Evolution of $V_{\rm thres}$ for each gate while $V_{\rm stress}$ is increased during the tuning procedure. The red dashed line indicates the target pinch-off voltage $V_{\rm target}=1.05~$V. The stressing on each gate is stopped when its pinch-off voltage becomes larger than $V_{\rm target}$. The coloring of the data points encodes the time evolution of the stress and pinch-off voltages of the gates during each iteration.}
\label{fig:Fig4}
\end{figure*}

Next we apply our findings and probe the capability to homogenize the pinch-off voltages $V_{\rm thres}^{\rm i}$ of a group of plunger gates $\rm P_{\rm i}$ with i~in~[1,4] in a quantum dot array. Fig~\ref{fig:Fig4}.a displays the device studied which has a geometry similar to linear quantum dot arrays in ref.~\cite{Zajac2016, Neyens2019, Noiri2022, Philips2022}. The pinch-off characteristics recorded prioir to the tuning sequence are depicted in the left panel of Fig~\ref{fig:Fig4}.c and show a spread $\Delta V_{\rm thres}=\max(V_{\rm thres})-\min(V_{\rm thres})$ of 153~mV. Employing increasing gate voltage stress we tune the individual plunger pinch-off voltages to a target value $V_{\rm target}=1.05$~V chosen before starting the tuning. Fig.~\ref{fig:Fig4}.b illustrates the procedure followed for the specific case of two gates. $V_{\rm stress}$ is gradually increased in steps~n. For each $V_{\rm stress}^{\rm n}$, the plunger gates are sequentially stressed, measured and parked $\Delta V_{\rm park}=50~$mV above their latest pinch-off voltage where they remain until the next stress voltage $V_{\rm stress}^{\rm n+1} = V_{\rm stress}^{\rm n} + \Delta V_{\rm stress}$ is selected. When a pinch-off voltage $V_{\rm thres}^{\rm i}$ crosses the target voltage $V_{\rm target}$ the corresponding plunger gate $\rm P_{\rm i}$ is henceforth not stressed anymore. A full automated round of this sequence finishes after all pinch-off voltages are larger than the target voltage. The complete procedure is repeated two times with a stress voltage resolution of $\Delta V_{\rm stress}=25$~mV taking approximately 9~hours in total. All applied stress voltages and measured pinch-off voltages are visualized in the panels of Fig.~\ref{fig:Fig4}.d. After each repetition a pinch-off characterization is performed with the resulting curves depicted in Fig.~\ref{fig:Fig4}.c. During the first round the pinch-off voltages shift towards the target voltage $V_{\rm target}$ (indicated by the red dashed line) finally spreading in a range of $\Delta V_{\rm thres}=86$~mV around it. This spread is further reduced by the following iteration reaching a final value of $\Delta V_{\rm thres}=20$~mV. Afterwards the plunger pinch-off characteristics are observed to remain stable at least for 20~minutes (see supplementary Fig.~S4).

To put this result into context, we compare the final spread of pinch-off voltages to the degree of uniformity needed to load an array of quantum dots with a single electron at each site using a single common gate voltage. This would require the potential fluctuations below the gates to be smaller than the average charging voltage $V_{\rm C}=E_{\rm C}/\alpha$ that is needed to alter the charge occupation, with $E_{\rm C}$ the charging energy and $\alpha$ the gate lever arm. This charging voltage typically ranges from 10~to~60~mV in devices similar to that under study~\cite{Zajac2016,Neyens2019,Lawrie2020,Takeda2021}. Assuming that pinch-off voltages constitute a witness of the intrinsic potential landscape in the quantum well, the final spread $\Delta V_{\rm thres}=20$~mV reached after electrical tuning  promises a path towards the homogenization of quantum dot potentials inside an array. Even smaller spreads might be achievable by decreasing the stress voltage resolution $\Delta V_{\rm stress}$. We envision that a similar method could be used to tune the potential underneath all plunger and all barrier gates simultaneously. It could allow to also equalize the inter-dot tunnel couplings and to reach an energy landscape similar to that in Fig.~\ref{fig:Fig1}.b.

At the same time, optimization of the automated procedure could lead to a significant increase of the tuning efficiency. Such an optimized procedure may be obtained by dividing the tuning into coarse and fine steps and exploring different stressing times and thereby could drastically reduce the tuning time. Additionally, utilizing a model to predict the effect of the next stress voltage, could further minimize the number of steps required to reach the target potentials and simultaneous tuning of multiple gates may be envisioned in larger quantum dot arrays.

Adapted tuning procedures may also be designed for scalable device architectures. In a crossbar gate architecture~\cite{Li2018,Borsoi2022}, one could envision to apply different stressing voltages on different sets of gates such that only close to the crossing points of these gates the combined electric field is strong enough to shift the intrinsic potential. This would allow parallel but individual stressing of selected sites in a row-by-row manner. Another degree of selectivity might be provided through  biasing of purposely isolated parts of the quantum well. Effectively, this would locally change the gates' reference potential and thereby locally alter the effect of the stressing voltages applied to them. Further work is needed to confirm the viability of these approaches.

Also, a better understanding of the underlying mechanism of the hysteresis would be valuable to exploit it most efficiently. A possible origin might be the trapping and detrapping of charge in or close to the dielectric capping layer caused by the application of stress voltages~\cite{Lu2011,Huang2014,Laroche2015,Su2017,Chou2018,Su2019}. For example, a positive stress voltage might enable the tunnelling of electrons from the quantum well or traps underneath non-stressed gates to traps underneath the stressed gate. These traps could be bound states in the non-oxidized part of the silicon capping layer or at its SiGe interface. They can be induced by charge defects in the gate oxide~\cite{Goetzberger1968} or emerge due to mechanical stress originating from the deposition of metallic gates~\cite{Thorbeck2015,Stein2021}. Also, charge trapping into and out of of unpassivated silicon and germanium dangling bonds~\cite{Poindexter1988,Lenahan1998,Stesmans2014}, charge trapping in the oxide itself mediated by leakage currents~\cite{Kerber2003,Pioro-Ladriere2005,Sze2006,Franco2014} or movement of mobile ions~\cite{Vanheusden1998} might be underlying the hysteresis. In all cases, when the gate voltage stress is removed the charges would be expected to be immobile at the device operation temperature and would cause local shifts in the intrinsic potential landscape observable as alterations in the pinch-off characteristics. This tunneling and trapping of charge also would be highly similar to the principle used to operate modern flash memories (based on electrically erasable programmable read only memories) which encode their stored information in pinch-off voltages and rely on gate stacks specifically engineered for that purpose~\cite{Hoffmann2004,Sze2006}. They could inspire new heterostructures and gate stacks with dedicated trapping layers further refining the tunability of the potential landscape using the gate voltage hysteresis.

In conclusion, we have presented a new method to increase the electrostatic potential uniformity in quantum dot devices electrically. We show that we can take advantage of hysteric shifts in gate voltage characteristics to deliberately tune pinch-off voltages across a wide range of more than 500~mV by applying stress voltage sequences. The resulting states remain stable on the time scale of hours. Utilizing our method, we have shifted and equalized the pinch-off voltages of four plunger gates in a linear quantum dot array to a predetermined target voltage. Although most of our results were obtained in Si/SiGe heterostructures other measurements indicate that the effect and method also can be used in other heterostructure materials like Ge/SiGe. Our work opens up a new path to increase uniformity in quantum dot based spin qubits. It may enable reducing overheads in tuning and control making the implementation of scalable architectures more feasible in practise.


\section*{Data availability}
The data and analysis supporting this work are openly available in a public Zenodo repository at \url{https://doi.org/10.5281/zenodo.7225478}~\cite{ZENODO_DATA}.

\section*{Acknowledgements}
We gratefully acknowledge D. Michalak, D. Degli-Esposti, M. Mehmandoost for sharing their expertise, their insights on the underlying physics and for their valuable advices. We also acknowledge S. Philips for his help on the Si/SiGe device designing. We  thank L.M.K Vandersypen for his feedback as well as all the members of the Veldhorst and Vandersypen group for stimulating discussions. We thank J. D. Mensingh and N. P. Alberts for their technical support with the experimental set-ups and S. L. de Snoo for software support.

We acknowledge support through an ERC Starting Grant and through an NWO projectruimte. This research was supported by the European Union’s Horizon 2020 research and innovation programme under the Grant Agreement No. 951852 (QLSI project). Research was sponsored by the Army Research Office (ARO) and was accomplished under Grant No. W911NF-17-1-0274. The views and conclusions contained in this document are those of the authors and should not be interpreted as representing the official policies, either expressed or implied, of the Army Research Office (ARO), or the U.S. Government. The U.S. Government is authorized to reproduce and distribute reprints for Government purposes notwithstanding any copyright notation herein.



\section*{Competing interest}
The authors declare no competing financial interest.

\bibliography{lib_manuscript}

\end{document}


\title{Supplementary Information - Electrical control of uniformity in quantum dot devices}

\author{Marcel Meyer}
\author{Corentin D\'{e}prez}
\author{Timo R. van Abswoude}
\author{Dingshan Liu}
\author{Chien-An Wang}
\affiliation{QuTech and Kavli Institute of Nanoscience, Delft University of Technology, PO Box 5046, 2600 GA Delft, The Netherlands}
\author{Saurabh Karwal}
\author{Stefan Oosterhout}
\affiliation{QuTech and Netherlands Organisation for Applied Scientific Research (TNO), Delft, The Netherlands}
\author{Francesco Borsoi}
\affiliation{QuTech and Kavli Institute of Nanoscience, Delft University of Technology, PO Box 5046, 2600 GA Delft, The Netherlands}
\author{Amir Sammak}
\affiliation{QuTech and Netherlands Organisation for Applied Scientific Research (TNO), Delft, The Netherlands}
\author{Nico W. Hendrickx}
\author{Giordanno Scappucci}
\author{Menno Veldhorst}
\affiliation{QuTech and Kavli Institute of Nanoscience, Delft University of Technology, PO Box 5046, 2600 GA Delft, The Netherlands}

\date{\today}

\begin{center}
\Huge{Supplementary Information}
\end{center}

\renewcommand\thefigure{S\arabic{figure}}
\renewcommand{\thetable}{T\arabic{table}}
\setcounter{figure}{0} 

\section{Material and Methods}
\label{section:material_and_methods}

\subsection{Materials and device fabrication}

The devices studied here are made from $^{28}$Si/SiGe heterostructures~\cite{Paquelet2021}. They are grown on top of a natural Si wafer and begin by a linearly graded Si$_{1-x}$Ge$_{x}$ wafer with $x$ varying from 0 to 0.3. A relaxed Si$_{0.7}$Ge$_{0.3}$ layer of 300~nm is then grown, followed by a 9~nm purified $^{28}\text{Si}$ layer (with 800 ppm purity) and another 30~nm thick relaxed Si$_{0.7}$Ge$_{0.3}$ layer. Finally, an approximately $1-2$~nm thin Si capping layer is deposited. The 2DEGs are contacted via phosphorus ion implantation. Overlapping Ti/Pd gate electrodes are deposited via electron beam evaporation. The different sets of gates are separated from each other and from the Si capping layer by 5~nm and 10~nm aluminum oxide layers, respectively, deposited through atomic layer deposition~\cite{Lawrie2020}.

\subsection{Experimental set-up}

All the measurements presented are dc-transport measurement performed at 4.2~K by dipping the devices directly in liquid helium. The gate voltages are swept using 18 bit precision digital-to-analog converters having a $\pm 4$~V range of applicable voltages. The current is measured via a current-to-voltage converter and a Keithley digital multimeter at an applied source-drain bias of $\abs{V_{\rm sd}}=100~\mu$V. The data acquisition, the application of the stress voltages and the successive pinch-off voltage measurements were performed automatically using a home-made Python program.   

\subsection{Experimental procedures}

Prior to any experiment, the group and individual pinch-off voltages $V_{\rm thres}$ of all gates forming a given conduction channel are measured. The group pinch-off voltage is measured by sweeping all gates corresponding to the channel under investigation simultaneously until a current of typically $200$ to $300$~pA is reached. The corresponding voltage $V_0$ is the voltage at which the gates not under study are parked most of the time. The individual pinch-off voltages are then measured by sweeping each gate voltage down and back up again starting from $V_0$. Such measurements allow to characterize the potential uniformity just after the cooldown and to potentially discard malfunctioning devices. The measurement of individual pinch-off voltages is repeated between experiments to study how the spread of the pinch-off voltages evolves and thereby the potential uniformity. To that end, first all gate voltages responsible for forming the conducting path are set to the same value.

In most experiments, before recording a ($V_{\rm stress}$,$V_{\rm thres}$) curve, a minimum and a maximum threshold voltage $V^{\rm min}_{\rm thres}$ and $V^{\rm max}_{\rm thres}$ are defined. Once $V_{\rm thres}<V^{\rm min}_{\rm thres}$ or $V_{\rm thres}>V^{\rm max}_{\rm thres}$ the sequence of stress voltages $V_{\rm stress}$ is reverted defining a reversal point $V_{\rm stress}^{\rm rev}$. Furthermore, we define minimum and maximum stress voltages $V^{\rm min}_{\rm stress}$ and $V^{\rm max}_{\rm stress}$ for each cycle and if $V_{\rm stress}\geq V^{\rm min}_{\rm stress}$ or $V_{\rm stress}\leq V^{\rm max}_{\rm stress}$ the sequence is also reversed defining a reversal point $V_{\rm stress}^{\rm rev}$ as well. Tables summarizing the reversal points for the different experiments can be found in supplementary table~\ref{table:Si_devices} and supplementary table~\ref{table:Ge_devices}.

We record the time evolution of a pinch-off voltage by continuously sweeping the corresponding gate voltage up and down around the pinch-off voltage. To that end the sweep direction is repeatedly reversed when the the measured current fulfills the condition $I>I_{\rm thres}+\Delta_{+}$ or $I<I_{\rm thres}+\Delta_{-}$, respectively, with $\Delta_{\pm}=30$~pA typically. This allows to extract the evolution of the pinch-off voltage with a time resolution of approximately $3$~s.

\section{Presentation of the Si/SiGe devices investigated}

\begin{figure}[h]
    \begin{centering}
    \includegraphics[width=\linewidth]{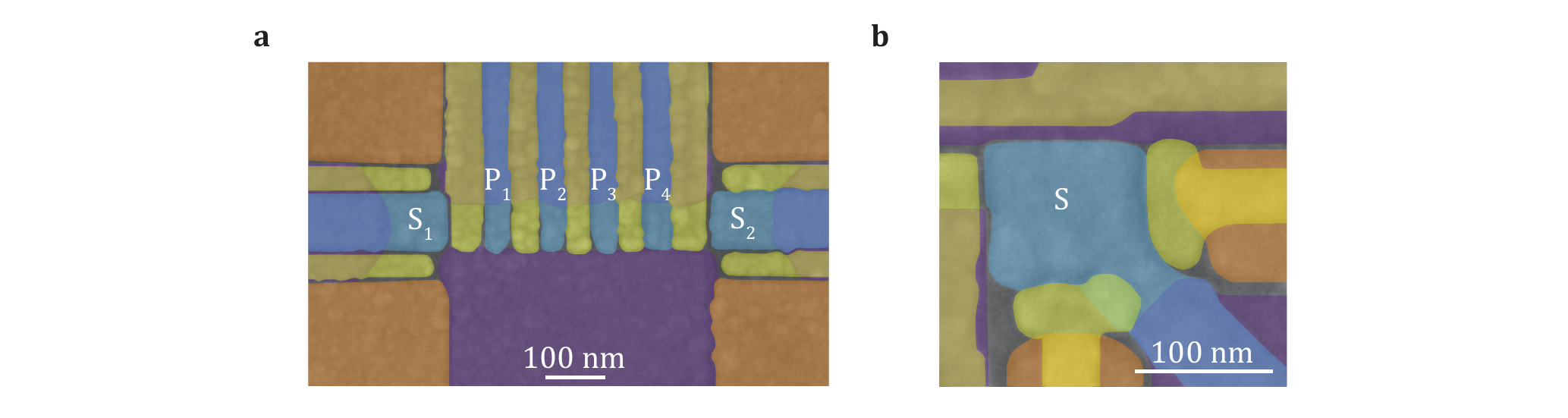}
    \end{centering}
    \caption{\textbf{Scanning electron micrograph of Si/SiGe devices studied.} \textbf{a}, Linear four quantum dot array, \textbf{b}, Single electron transistor at the edge of a 3$\times$3 quantum dot array. The plunger gates are colored in yellow, the barrier gates in blue, the accumulation gates in orange and the screening gates in violet. In \textbf{a}, the plunger gates belonging to the linear channel are labelled $\rm P_{\rm i}$ whereas that of the charge sensors are labelled $\rm S_{\rm i}$. In \textbf{b}, the plunger gate of the sensor used during the experiments is labelled $\rm S$. }
    \label{fig:Sup_SEM_Si}
\end{figure}

Here, we present the Si/SiGe quantum dot devices investigated in this work. Fig~\ref{fig:Sup_SEM_Si} shows typical scanning electron micrographs of the two types of devices. They are both composed of screening, accumulation/plunger and barrier gates which are deposited in that order. The screening gates are usually kept close to 0~V to prevent the formation of conducting channels at undesired locations. The devices labeled A and B in the main text are nominally identical to the one presented in Fig~\ref{fig:Sup_SEM_Si}.a. It is designed to form a linear array of four quantum dots with two larger quantum dots at the ends to be used as charge sensors. The devices named C and D are of the type displayed in Fig~\ref{fig:Sup_SEM_Si}.b which shows a single quantum dot aimed to be a single electron transistor. It is located at the corner of a larger 3$\times$3 quantum dot array (not shown here).

\section{Reproducibility of the ($V_{\rm thres}$,$V_{\rm stress}$) hysteresis cycles}

\begin{figure}[h]
    \begin{centering}
    \includegraphics[width=\linewidth]{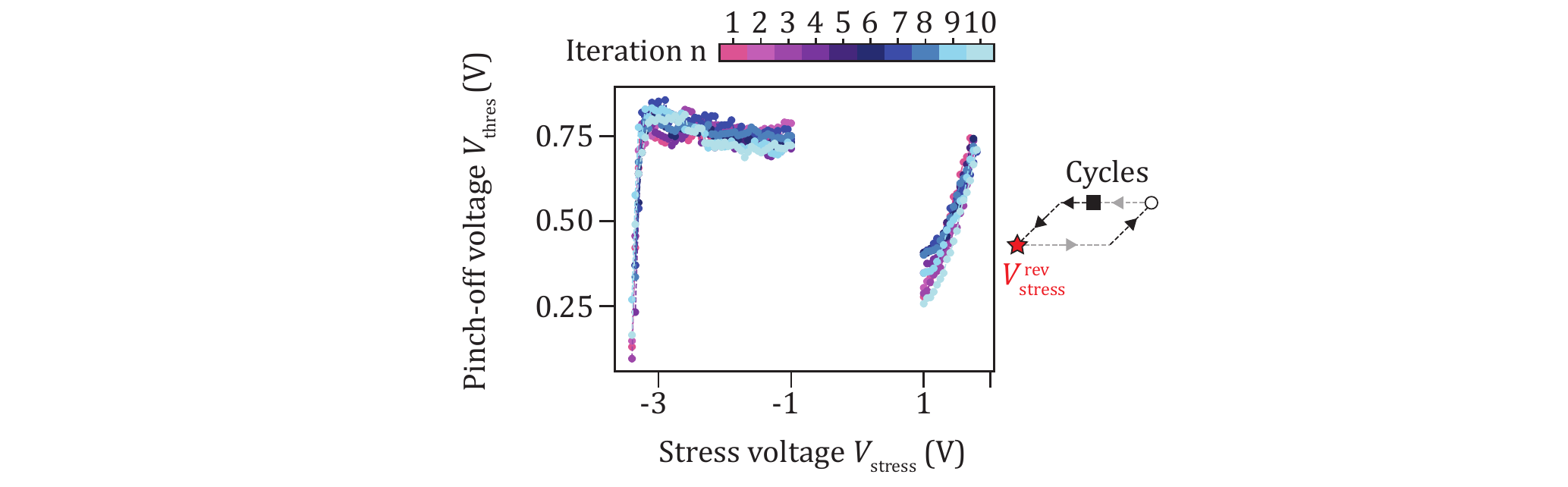}
    \end{centering}
    \caption{\textbf{Overlap of multiple hysteresis cycles.} Evolution of $V_{\rm thres}$ as a function of $V_{\rm stress}$ for 10 successive cycles obtained for gate~S in device~D. A schematic of the stress voltage sequence applied during one iteration is sketched on the right. The square and the circle mark the starting point and the ending point of the cycles, respectively. The star indicates the point where the stress voltage direction is reversed. The $V_{\rm thres}$ plateaus are not measured or only partially (grey dashed lines). All the curves collapse onto each other showing a remarkable reproducibility.}
    \label{fig:Sup_10times}
\end{figure}

To provide further evidence of the reproducibility of the hysteresis cycles, we perform an experiment where the same sequences of decreasing and then increasing stress voltages are repeated ten times for gate~S in device~D. To reduce the measurement time, we focus mostly on the voltage range where $V_{\rm thres}$ shows a strong evolution with $V_{\rm stress}$. The results are displayed in Fig.~\ref{fig:Sup_10times}.

We recognize the left and right flanks of the hysteresis cycles as well as the end of the $V_{\rm thres}$ plateaus for decreasing $V_{\rm stress}$. Remarkably, the data obtained for the ten iterations collapse onto single curves both for increasing and decreasing $V_{\rm stress}$. This further illustrates the high reproducibility of the $(V_{\rm thres}$,$V_{\rm stress})$ hysteresis cycle that can be achieved.

\section{Pinch-off voltages hysteresis in Ge/SiGe}
\label{Germanium_devices}

\begin{figure}[h]
    \begin{centering}
    \includegraphics[width=1\linewidth]{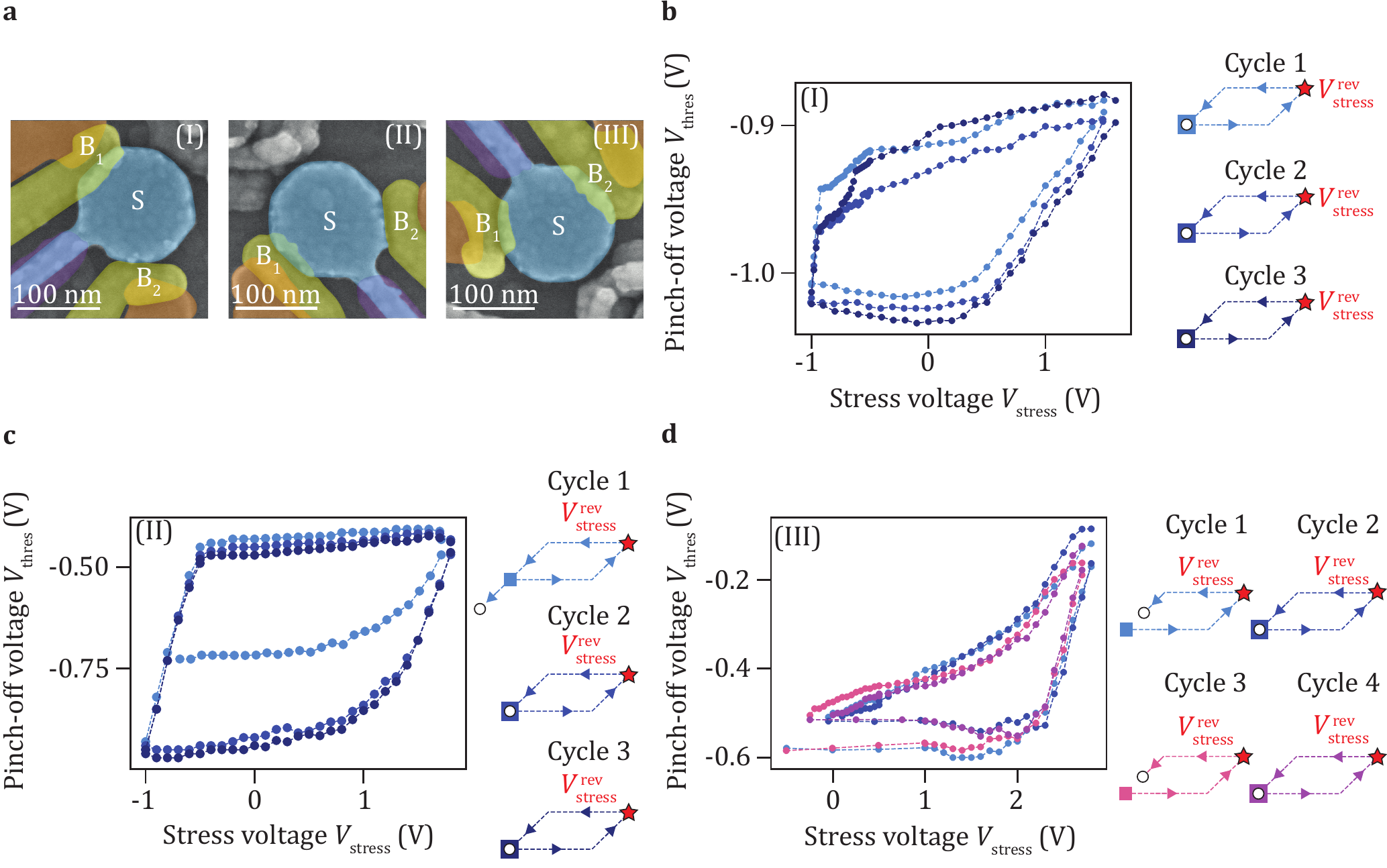}
    \end{centering}
    \caption{\textbf{Hysteresis of $V_{\rm thres}$ in Ge/SiGe single hole transistors. a}, Scanning electron micrograph of the SHTs studied. The plunger gates are colored in blue, the barrier gates in yellow, the ohmic contacts in orange and the screening gates in violet. The plunger gates of the SHTs used during the experiments are labelled $\rm S$ while the two barrier gates are labelled as $\rm B_{\rm1,2}$. \textbf{b}, \textbf{c}, \textbf{d} Evolution of $V_{\rm thres}$ as a function of $V_{\rm stress}$ for successive stress voltage cycles applied simultaneously on the barrier gates and the plunger gate forming the SHT. The stress voltage cycles are depicted schematically on the right side. The square and the circle mark the starting point and the ending point of the cycles, respectively. The stress voltage $V_{\rm stress}^{\rm rev}$ upon which the stress voltage sequence is reversed is indicated by a star. For each SHT, the different data points overlap and form a hysteresis loop. In \textbf{d}, the device was kept idle for 5 hours between cycle 2 and cycle 3. This lead to a voltage difference $\Delta V_{\rm thres}\approx-80$~mV between the last point of cycle 2 and the first point of cycle 3 similar to the time evolution discussed Fig.~3. The data is taken at an electron temperature of approximately 140~mK~\cite{Borsoi2022}.}
    \label{fig:Sup_Germanium}
\end{figure}

The hysteretic behavior of the pinch-off voltages and its dependence on the previous stress voltages applied is not exclusive to Si/SiGe heterostructures. The same effect can  be observed in Ge/SiGe heterostructures. We perform experiments similar to those discussed in the main text in germanium single hole transistor (SHT) structures that are presented in Fig.~\ref{fig:Sup_Germanium}.a. Note that these SHTs are part of a larger device presented in ref.~\cite{Borsoi2022}.

The corresponding device is made from a strained Ge/SiGe heterostructures grown by chemical vapor deposition. Starting from a natural Si wafer, a 1.6~$\mu$m thick relaxed Ge layer is grown, followed by a 1$~\mu$m reverse graded Si$_{1-x}$Ge$_x$ ($x$ going from 1 to 0.8) layer, a 500~nm relaxed Si$_{0.2}$Ge$_{0.8}$ layer, a 16~nm Ge quantum well under compressive stress, another 55~nm Si$_{0.2}$Ge$_{0.8}$ spacer layer and a $<1$~nm thick Si capping layer~\cite{Sammak2019, Lodari2021}. The quantum well is contacted via 30-nm platinium contacts evaporated and diffused after etching of the oxidized Si capping layer~\cite{Hendrickx2018}. Aluminum oxide layers of 7, 5, and 5~nm thickness grown by atomic layer deposition precede the deposition of overlapping Ti/Pd electrodes with thicknesses of 3/17, 3/27, 3/27~nm forming three different gate layers on top of the heterostructure~\cite{Lawrie2020}. 

We study the devices by applying a common gate voltage to the two barrier gates and the plunger gate defining the SHT such that a conductive channel is formed between the ohmic contacts. Fig.~\ref{fig:Sup_Germanium}.b-d show typical hysteresis cycles obtained by measuring the evolution of the pinch-off voltage as function of the stress voltage applied on the three gates for each SHT. $V_{\rm stress}$ is first increased and then decreased in each measurement cycle contrary to the sequence followed in Fig.~2.  These measurements are performed at base temperature of a dilution refrigerator and an estimated electron temperature of approximately 140~mK~\cite{Borsoi2022} and the pinch-off voltage was defined as the voltage at which the current reaches $I_{\rm thres}=500~$pA at an source drain bias of $\abs{V_{\rm sd}}=100~\upmu$V.

Overall, we observe similar features to those observed in the Si/SiGe devices i.e. overlapping hysteresis cycles with a tunable voltage range of a few hundred millivolts. These measurements highlight that the hysteresis of the pinch-off voltages is observable in multiple semiconductor heterostructures and that the tuning method presented in this work may be used in different materials as well.

\section{Stability of the pinch-off characteristics after tuning them using the hysteretic behaviour}

\begin{figure}[h]
    \begin{centering}
    \includegraphics[width=\linewidth]{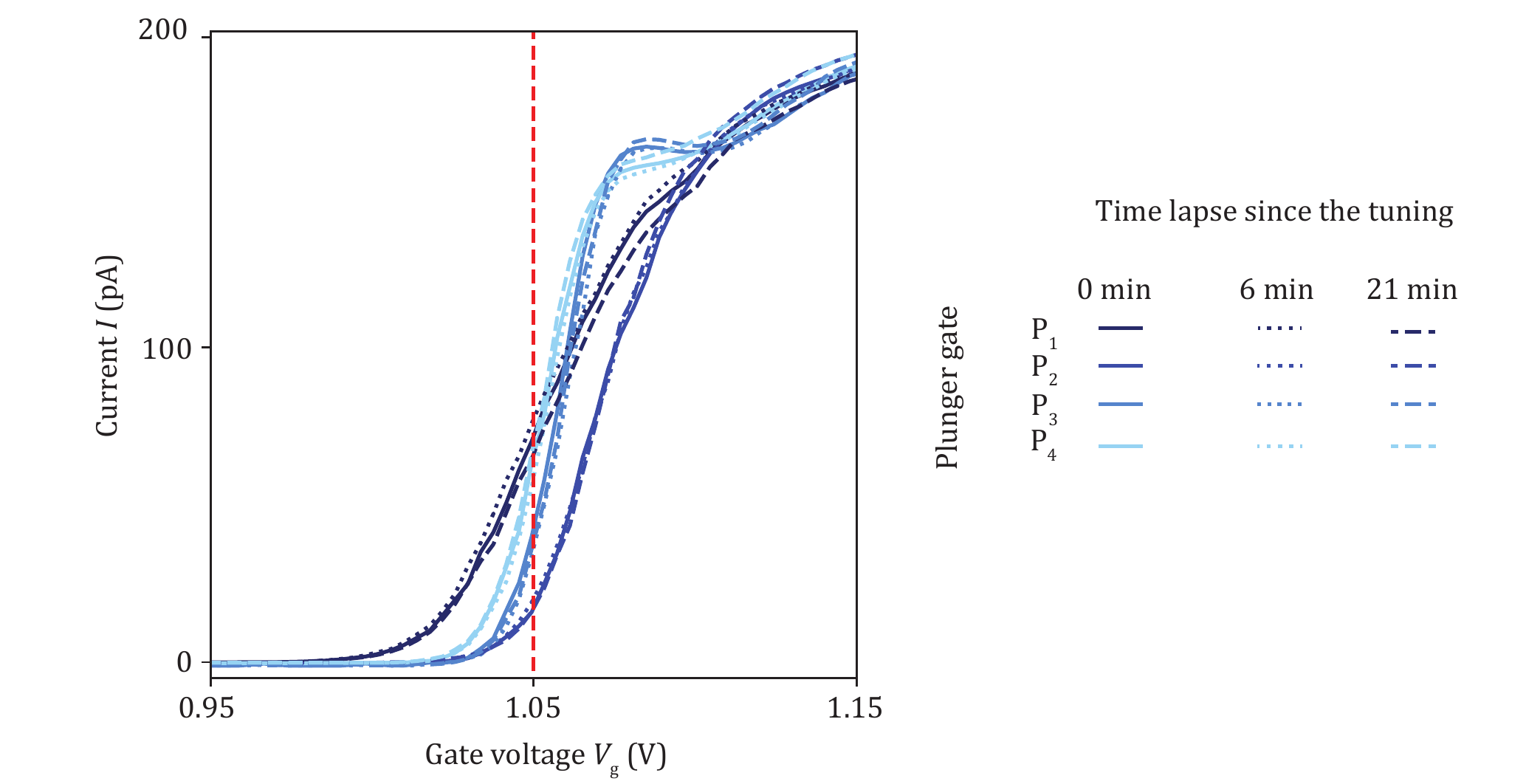}
    \end{centering}
    \caption{\textbf{Stability of the plunger gate pinch-off characteristics after tuning them to the target voltage.}  Pinch-off characteristics measured just after (plain lines), 6~minutes (dotted lines) and 21~minutes (dashed lines) after the homogeneization procedure described in the main text using the hysteretic shift. The red dashed line marks the target voltage of the tuning.}
    \label{fig:Sup_stability}
\end{figure}

Here, we discuss the stability of the pinch-off voltages after reduction of their spread using the protocol presented in Fig.~4. Fig.~\ref{fig:Sup_stability} shows the pinch-off characteristics right after, 6, and 21~minutes after the tuning.

For each plunger gate considered, the three plots virtually overlap perfectly suggesting a high degree of stability. This also is in agreement with the absence of variations observed in the time stability measured after application of increasing stress voltages in Fig.~3. It supports our choice of using increasing stress voltages in the tuning procedure presented.

\section{Samples used for each measurements and reversal points}

Table~\ref{table:Si_devices} and Table~\ref{table:Ge_devices} provide an overview of the different samples underlying the figures in this work, their gate design, the gates that were swept, and the reversal points $V_{\rm stress}^{\rm rev}$ after which the stress voltage sequence was reversed if applicable. The gate designs and gate names mentioned in Table~\ref{table:Si_devices} can be found in supplementary Fig.~\ref{fig:Sup_SEM_Si}. Supplementary Fig.~\ref{fig:Sup_Germanium} shows gate designs and gate names for the SHTs referred to in Table~\ref{table:Ge_devices}.

\begin{table}[h!]
\begin{center}
\begin{tabular}{ |c|c|c|c|c|c| } 
 \hline
Figure & Device & Type of  & Gate(s)  & Cycle number & Reversal/end point(s) $V_{\rm stress}^{\rm rev/end}$  \\ 
label & measured &  device & swept &  & of the stress voltage sequence (in V) \\ \hline \hline
Fig~1 & A & Linear array & P$_1$, P$_2$, P$_3$, P$_4$ & N.A. & N.A. \\ \hline 
Fig~2.b & B & Linear array & S$_1$ & 1 & -3.7 / 1.7 \\ \hline
Fig~2.c & A & Linear array & P$_1$ & 1 & -2.4 / 2.4 \\ 
 &  & & & 2 & -2.4 / 2.5 \\ 
 &  &  &  & 3 & -2.3 / 2.6 \\ 
 &  & &  & 4 & -2.3 / 2.7 \\ 
 & &  & & 5 & -2.6 / 2.1 \\ \hline
Fig~3 (blue, orange, & C & $3\times 3$ array & S & N.A. & N.A.\\ 
 violet curve) &  &  &  &  & \\ \hline
Fig~3 (pink curve) & D & $3\times 3$ array & S & N.A. & N.A.\\\hline
Fig~4  & A & Linear array & P$_1$, P$_2$, P$_3$, P$_4$  & N.A. & N.A.\\\hline
Fig~\ref{fig:Sup_10times}  & D & $3\times 3$ & S & 1 & -3.4 / 1.7 \\
 &  & & & 2 & -3.4 / 1.75 \\ 
 &  & & & 3 & -3.4 / 1.75 \\ 
 &  & & & 4 & -3.35 / 1.75 \\ 
 &  & & & 5 & -3.35 / 1.7 \\ 
 &  & & & 6 & -3.35 / 1.75 \\ 
 &  & & & 7 & -3.35 / 1.8 \\ 
 &  & & & 8 & -3.35 / 1.75 \\ 
 &  & & & 9 & -3.4 / 1.75 \\
 &  & & & 10 & -3.4 / 1.8 \\ \hline
Fig~\ref{fig:Sup_stability}  & A & Linear array & P$_1$, P$_2$, P$_3$, P$_4$  & N.A. & N.A.\\\hline

\end{tabular}
\end{center}

\caption{\textbf{Summary table for the Si/SiGe devices}}
\label{table:Si_devices}
\end{table}

\begin{table}[h!]
\begin{center}
\begin{tabular}{ |c|c|c|c|c|c| } 
 \hline
Figure & SHT & Type of  & Gates  & Cycle number & Reversal/end point(s) $V_{\rm stress}^{\rm rev/end}$  \\ 
label & measured &  device & swept &  &  of the stress voltage sequence (in V) \\ \hline \hline
Fig~\ref{fig:Sup_Germanium}.b & I & $4\times 4$ array & S, $\rm B_{1}$, $\rm B_{2}$ & 1 & 1.5 / -1.0 \\ 
 &  &  & (simultaneously) & 2 & 1.492 / -1.0 \\ 
 &  &  &  & 3 & -1.6 / -1.0 \\ \hline 
Fig~\ref{fig:Sup_Germanium}.c & II & $4\times 4$ array & S, $\rm B_{1}$, $\rm B_{2}$ & 1 & 1.714 / -1.0 \\ 
 &  &  & (simultaneously) & 2 & 1.8 / -1.0 \\ 
 &  &  &  & 3 & 1.8 / -1.0 \\ \hline 
Fig~\ref{fig:Sup_Germanium}.d & III & $4\times 4$ array & S, $\rm B_{1}$, $\rm B_{2}$ & 1 & 2.8 / 0.05 \\ 
 & &  & (simultaneously) & 2 & 2.8 / 0.1 \\  
 &  &  &  & 3 & 1.7 / -0.25 \\ 
&  &  &  & 4 & 2.7 / 0.0 \\ \hline 

\end{tabular}
\end{center}

\caption{\textbf{Summary table for Ge/SiGe devices}}
\label{table:Ge_devices}
\end{table}

\bibliography{lib_sup}